\documentstyle[10pt,epsf,dp_delphititle]{dp_delphi}
\makeindex
\def\DpPaperGroup{EP}
\def\DpPaperRef{98--171}
\def\DpDate{{28 October 1998}}
\def\DpAuthors{DELPHI Collaboration}
\def\DpTitle{{A Search for Heavy Stable and Long-Lived Squarks and Sleptons in 
\boldmath e$^{+}$e$^{-}$ Collisions at Energies from 130 to 183 GeV}}
\def\DpSubmit{(Submitted to Physics Letters B)}

\begin{document}
\makeatletter
\newcount\@tempcntc
\def\@citex[#1]#2{\if@filesw\immediate\write\@auxout{\string\citation{#2}}\fi
  \@tempcnta\z@\@tempcntb\m@ne\def\@citea{}\@cite{\@for\@citeb:=#2\do
    {\@ifundefined
       {b@\@citeb}{\@citeo\@tempcntb\m@ne\@citea\def\@citea{,}{\bf ?}\@warning
       {Citation `\@citeb' on page \thepage \space undefined}}%
    {\setbox\z@\hbox{\global\@tempcntc0\csname b@\@citeb\endcsname\relax}%
     \ifnum\@tempcntc=\z@ \@citeo\@tempcntb\m@ne
       \@citea\def\@citea{,}\hbox{\csname b@\@citeb\endcsname}%
     \else
      \advance\@tempcntb\@ne
      \ifnum\@tempcntb=\@tempcntc
      \else\advance\@tempcntb\m@ne\@citeo
      \@tempcnta\@tempcntc\@tempcntb\@tempcntc\fi\fi}}\@citeo}{#1}}
\def\@citeo{\ifnum\@tempcnta>\@tempcntb\else\@citea\def\@citea{,}%
  \ifnum\@tempcnta=\@tempcntb\the\@tempcnta\else
   {\advance\@tempcnta\@ne\ifnum\@tempcnta=\@tempcntb \else \def\@citea{--}\fi
    \advance\@tempcnta\m@ne\the\@tempcnta\@citea\the\@tempcntb}\fi\fi}
 
\makeatother
\begin{titlepage}
\pagenumbering{roman}
\CERNpreprint{\DpPaperGroup}{\DpPaperRef} 
\date{{\small\DpDate}} 
\title{\DpTitle} 
\address{\DpAuthors} 
\begin{shortabs} 
\noindent
%
\noindent

A search for stable and long-lived heavy charged particles 
used the data taken by the DELPHI experiment 
at energies from 130 to 183 GeV.
The Cherenkov light detected in the Ring Imaging Cherenkov Detector and the 
ionization loss 
measured in the Time Projection Chamber identify heavy particles from 
masses of 2 to nearly 89 GeV/c$^2$.
Upper limits are given 
on the production cross-section and masses of sleptons,
free squarks with a charge of $q=\pm \frac{2}{3}e$ and hadronizing squarks.
\end{shortabs}
\vfill
\begin{center}
\DpSubmit \ 
\end{center}
\vfill
\clearpage
\headsep 10.0pt
\addtolength{\textheight}{10mm}
\addtolength{\footskip}{-5mm}
\begingroup
%
\newcommand{\DpName}[2]{\hbox{#1$^{\ref{#2}}$},\hfill}
\newcommand{\DpNameTwo}[3]{\hbox{#1$^{\ref{#2},\ref{#3}}$},\hfill}
\newcommand{\DpNameThree}[4]{\hbox{#1$^{\ref{#2},\ref{#3},\ref{#4}}$},\hfill}
\newskip\Bigfill \Bigfill = 0pt plus 1000fill
\newcommand{\DpNameLast}[2]{\hbox{#1$^{\ref{#2}}$}\hspace{\Bigfill}}
%
\footnotesize
\noindent
\DpName{P.Abreu}{LIP}
\DpName{W.Adam}{VIENNA}
\DpName{T.Adye}{RAL}
\DpName{P.Adzic}{DEMOKRITOS}
\DpName{T.Aldeweireld}{AIM}
\DpName{G.D.Alekseev}{JINR}
\DpName{R.Alemany}{VALENCIA}
\DpName{T.Allmendinger}{KARLSRUHE}
\DpName{P.P.Allport}{LIVERPOOL}
\DpName{S.Almehed}{LUND}
% ...
\normalsize
\endgroup
\titlefoot{Department of Physics and Astronomy, Iowa State
     University, Ames IA 50011-3160, USA
    \label{AMES}}
\titlefoot{Physics Department, Univ. Instelling Antwerpen,
     Universiteitsplein 1, BE-2610 Wilrijk, Belgium \\
     \indent~~and IIHE, ULB-VUB,
     Pleinlaan 2, BE-1050 Brussels, Belgium \\
     \indent~~and Facult\'e des Sciences,
     Univ. de l'Etat Mons, Av. Maistriau 19, BE-7000 Mons, Belgium
    \label{AIM}}
\titlefoot{Physics Laboratory, University of Athens, Solonos Str.
     104, GR-10680 Athens, Greece
    \label{ATHENS}}
\titlefoot{Department of Physics, University of Bergen,
     All\'egaten 55, NO-5007 Bergen, Norway
    \label{BERGEN}}
\titlefoot{Dipartimento di Fisica, Universit\`a di Bologna and INFN,
     Via Irnerio 46, IT-40126 Bologna, Italy
    \label{BOLOGNA}}
\titlefoot{Centro Brasileiro de Pesquisas F\'{\i}sicas, rua Xavier Sigaud 150,
     BR-22290 Rio de Janeiro, Brazil \\
     \indent~~and Depto. de F\'{\i}sica, Pont. Univ. Cat\'olica,
     C.P. 38071 BR-22453 Rio de Janeiro, Brazil \\
     \indent~~and Inst. de F\'{\i}sica, Univ. Estadual do Rio de Janeiro,
     rua S\~{a}o Francisco Xavier 524, Rio de Janeiro, Brazil
    \label{BRASIL}}
\titlefoot{Comenius University, Faculty of Mathematics and Physics,
     Mlynska Dolina, SK-84215 Bratislava, Slovakia
    \label{BRATISLAVA}}
\titlefoot{Coll\`ege de France, Lab. de Physique Corpusculaire, IN2P3-CNRS,
     FR-75231 Paris Cedex 05, France
    \label{CDF}}
\titlefoot{CERN, CH-1211 Geneva 23, Switzerland
    \label{CERN}}
\titlefoot{Institut de Recherches Subatomiques, IN2P3 - CNRS/ULP - BP20,
     FR-67037 Strasbourg Cedex, France
    \label{CRN}}
\titlefoot{Institute of Nuclear Physics, N.C.S.R. Demokritos,
     P.O. Box 60228, GR-15310 Athens, Greece
    \label{DEMOKRITOS}}
\titlefoot{FZU, Inst. of Phys. of the C.A.S. High Energy Physics Division,
     Na Slovance 2, CZ-180 40, Praha 8, Czech Republic
    \label{FZU}}
\titlefoot{Dipartimento di Fisica, Universit\`a di Genova and INFN,
     Via Dodecaneso 33, IT-16146 Genova, Italy
    \label{GENOVA}}
\titlefoot{Institut des Sciences Nucl\'eaires, IN2P3-CNRS, Universit\'e
     de Grenoble 1, FR-38026 Grenoble Cedex, France
    \label{GRENOBLE}}
\titlefoot{Helsinki Institute of Physics, HIP,
     P.O. Box 9, FI-00014 Helsinki, Finland
    \label{HELSINKI}}
\titlefoot{Joint Institute for Nuclear Research, Dubna, Head Post
     Office, P.O. Box 79, RU-101 000 Moscow, Russian Federation
    \label{JINR}}
\titlefoot{Institut f\"ur Experimentelle Kernphysik,
     Universit\"at Karlsruhe, Postfach 6980, DE-76128 Karlsruhe,
     Germany
    \label{KARLSRUHE}}
\titlefoot{Institute of Nuclear Physics and University of Mining and Metalurgy,
     Ul. Kawiory 26a, PL-30055 Krakow, Poland
    \label{KRAKOW}}
\titlefoot{Universit\'e de Paris-Sud, Lab. de l'Acc\'el\'erateur
     Lin\'eaire, IN2P3-CNRS, B\^{a}t. 200, FR-91405 Orsay Cedex, France
    \label{LAL}}
\titlefoot{School of Physics and Chemistry, University of Lancaster,
     Lancaster LA1 4YB, UK
    \label{LANCASTER}}
\titlefoot{LIP, IST, FCUL - Av. Elias Garcia, 14-$1^{o}$,
     PT-1000 Lisboa Codex, Portugal
    \label{LIP}}
\titlefoot{Department of Physics, University of Liverpool, P.O.
     Box 147, Liverpool L69 3BX, UK
    \label{LIVERPOOL}}
\titlefoot{LPNHE, IN2P3-CNRS, Univ.~Paris VI et VII, Tour 33 (RdC),
     4 place Jussieu, FR-75252 Paris Cedex 05, France
    \label{LPNHE}}
\titlefoot{Department of Physics, University of Lund,
     S\"olvegatan 14, SE-223 63 Lund, Sweden
    \label{LUND}}
\titlefoot{Universit\'e Claude Bernard de Lyon, IPNL, IN2P3-CNRS,
     FR-69622 Villeurbanne Cedex, France
    \label{LYON}}
\titlefoot{Univ. d'Aix - Marseille II - CPP, IN2P3-CNRS,
     FR-13288 Marseille Cedex 09, France
    \label{MARSEILLE}}
\titlefoot{Dipartimento di Fisica, Universit\`a di Milano and INFN,
     Via Celoria 16, IT-20133 Milan, Italy
    \label{MILANO}}
\titlefoot{Niels Bohr Institute, Blegdamsvej 17,
     DK-2100 Copenhagen {\O}, Denmark
    \label{NBI}}
\titlefoot{NC, Nuclear Centre of MFF, Charles University, Areal MFF,
     V Holesovickach 2, CZ-180 00, Praha 8, Czech Republic
    \label{NC}}
\titlefoot{NIKHEF, Postbus 41882, NL-1009 DB
     Amsterdam, The Netherlands
    \label{NIKHEF}}
\titlefoot{National Technical University, Physics Department,
     Zografou Campus, GR-15773 Athens, Greece
    \label{NTU-ATHENS}}
\titlefoot{Physics Department, University of Oslo, Blindern,
     NO-1000 Oslo 3, Norway
    \label{OSLO}}
\titlefoot{Dpto. Fisica, Univ. Oviedo, Avda. Calvo Sotelo
     s/n, ES-33007 Oviedo, Spain
    \label{OVIEDO}}
\titlefoot{Department of Physics, University of Oxford,
     Keble Road, Oxford OX1 3RH, UK
    \label{OXFORD}}
\titlefoot{Dipartimento di Fisica, Universit\`a di Padova and
     INFN, Via Marzolo 8, IT-35131 Padua, Italy
    \label{PADOVA}}
\titlefoot{Rutherford Appleton Laboratory, Chilton, Didcot
     OX11 OQX, UK
    \label{RAL}}
\titlefoot{Dipartimento di Fisica, Universit\`a di Roma II and
     INFN, Tor Vergata, IT-00173 Rome, Italy
    \label{ROMA2}}
\titlefoot{Dipartimento di Fisica, Universit\`a di Roma III and
     INFN, Via della Vasca Navale 84, IT-00146 Rome, Italy
    \label{ROMA3}}
\titlefoot{DAPNIA/Service de Physique des Particules,
     CEA-Saclay, FR-91191 Gif-sur-Yvette Cedex, France
    \label{SACLAY}}
\titlefoot{Instituto de Fisica de Cantabria (CSIC-UC), Avda.
     los Castros s/n, ES-39006 Santander, Spain
    \label{SANTANDER}}
\titlefoot{Dipartimento di Fisica, Universit\`a degli Studi di Roma
     La Sapienza, Piazzale Aldo Moro 2, IT-00185 Rome, Italy
    \label{SAPIENZA}}
\titlefoot{Inst. for High Energy Physics, Serpukov
     P.O. Box 35, Protvino, (Moscow Region), Russian Federation
    \label{SERPUKHOV}}
\titlefoot{J. Stefan Institute, Jamova 39, SI-1000 Ljubljana, Slovenia
     and Department of Astroparticle Physics, School of\\
     \indent~~Environmental Sciences, Kostanjeviska 16a, Nova Gorica,
     SI-5000 Slovenia, \\
     \indent~~and Department of Physics, University of Ljubljana,
     SI-1000 Ljubljana, Slovenia
    \label{SLOVENIJA}}
\titlefoot{Fysikum, Stockholm University,
     Box 6730, SE-113 85 Stockholm, Sweden
    \label{STOCKHOLM}}
\titlefoot{Dipartimento di Fisica Sperimentale, Universit\`a di
     Torino and INFN, Via P. Giuria 1, IT-10125 Turin, Italy
    \label{TORINO}}
\titlefoot{Dipartimento di Fisica, Universit\`a di Trieste and
     INFN, Via A. Valerio 2, IT-34127 Trieste, Italy \\
     \indent~~and Istituto di Fisica, Universit\`a di Udine,
     IT-33100 Udine, Italy
    \label{TU}}
\titlefoot{Univ. Federal do Rio de Janeiro, C.P. 68528
     Cidade Univ., Ilha do Fund\~ao
     BR-21945-970 Rio de Janeiro, Brazil
    \label{UFRJ}}
\titlefoot{Department of Radiation Sciences, University of
     Uppsala, P.O. Box 535, SE-751 21 Uppsala, Sweden
    \label{UPPSALA}}
\titlefoot{IFIC, Valencia-CSIC, and D.F.A.M.N., U. de Valencia,
     Avda. Dr. Moliner 50, ES-46100 Burjassot (Valencia), Spain
    \label{VALENCIA}}
\titlefoot{Institut f\"ur Hochenergiephysik, \"Osterr. Akad.
     d. Wissensch., Nikolsdorfergasse 18, AT-1050 Vienna, Austria
    \label{VIENNA}}
\titlefoot{Inst. Nuclear Studies and University of Warsaw, Ul.
     Hoza 69, PL-00681 Warsaw, Poland
    \label{WARSZAWA}}
\titlefoot{Fachbereich Physik, University of Wuppertal, Postfach
     100 127, DE-42097 Wuppertal, Germany
    \label{WUPPERTAL}}
\titlefoot{On leave of absence from IHEP Serpukhov
    \label{MILAN-SERPOU}}
\titlefoot{Now at University of Florida
    \label{FLORIDA}}
\addtolength{\textheight}{-10mm}
\addtolength{\footskip}{5mm}
\clearpage
\headsep 30.0pt
\end{titlepage}
%
\pagestyle{heading} 
\pagenumbering{arabic} 
\renewcommand{\thefootnote}{\fnsymbol{footnote}} 
\setcounter{footnote}{0} %
\large
%
\section{Introduction}

A search for stable and long-lived\footnote{Throughout the paper 
stable particles include long-lived particles decaying outside the detector.}
heavy charged particles 
in {\it all} final states is reported using the data taken by the DELPHI 
experiment at energies from 130 to 183 GeV. These results
extend those reported in \cite{ref:hdelphi} by including the 
130-136 and 183 GeV data taken in 1997. The other LEP experiments have searched 
for stable and long-lived heavy charged particles in low multiplicity final 
states \cite{ref:hlep}.

In most models of Supersymmetry (SUSY) the supersymmetric partners of standard
particles are unstable and have short lifetimes, except the lightest  supersymmetric
particle (LSP) which could be neutral and stable.
In most of the searches it is therefore assumed that the 
supersymmetric particles decay promptly. 
However, it is possible that a stable or
long-lived heavy charged SUSY-particle exists.
In the Minimal Supersymmetric Standard Model (MSSM) 
with the neutralino as the LSP\cite{ref:susy},
if the mass difference between the chargino and neutralino is small
the chargino can have a 
sufficiently long lifetime to be observed as stable in the detector.
In the MSSM with a very small amount of R-parity violation 
the LSP can be a charged slepton or squark and decay with a long lifetime 
into Standard Model particles \cite{ref:rpar}.

In gauge mediated supersymmetric models the gravitino
is the LSP and the next to lightest supersymmetric particle 
(NLSP) could have a long lifetime in a very natural way 
for large values of the SUSY-breaking scale \cite{ref:slepton}.
This is possible for sleptons, for example when the stau is the NLSP. 
In certain 
variations of the minimal model the squark 
can be the NLSP and become long-lived \cite{ref:giu}.

Other SUSY and non-SUSY models predict 
stable and long-lived heavy charged leptons, quarks 
and hadrons not present in the Standard Model. 
Free (s)quarks  might even exist \cite{ref:free}.

The published analyses from DELPHI \cite{ref:hdelphi} and the other LEP 
experiments \cite{ref:hlep} covered masses, $m$, above 45 GeV/c$^2$.
The present analysis has been further optimized for squarks 
and extended down to masses of 2 GeV/c$^2$.
This extension is important for the stable and long-lived squark search.
Stable long-lived free squarks of charge $\pm \frac{2}{3}e$ were excluded 
by the data taken at the Z$^0$ peak \cite{ref:sdelphi}. However, the upper limits 
on the production cross-section of squarks, where the squark dresses up and 
becomes a charged or neutral shadron in a hadronization or 
fragmentation process,
are worse than those of free squarks. In particular,
hadronizing stop and sbottom quarks with 
so-called typical mixing and down-type right-handed squarks 
are not ruled out in the mass region from $\sim$15 to 45 GeV/c$^2$
due to the small production cross-section at Z$^0$ energies.

Limits on the production cross-section and masses will be given 
for stable and long-lived sleptons, charginos,
free (not hadronizing) squarks of charge q = $\pm \frac{2}{3}e$ and hadronizing squarks 
(q = $\pm \frac{1}{3}e$ or $\pm \frac{2}{3}e$) forming shadrons. No search 
is made for free squarks of charge q = $\pm \frac{1}{3}e$, because 
the tracking system is not sensitive enough to record the ionization of
these particles. 

A dedicated simulation
program was used for the hadronization of squarks.
It is assumed that the sleptons, charginos, free squarks and shadrons 
decay outside the tracking volume of the detector,
which extends to a typical radius of 1.5 m. It is further assumed 
that these particles 
do not interact more strongly 
than ordinary matter particles (protons or electrons)
and reach the main tracking device.

 Heavy stable particles are selected by looking for high momentum
charged particles with either anomalous ionization loss dE/dx 
measured in the Time Projection Chamber (TPC), or the absence of 
Cherenkov light in the gas and liquid radiators of 
the Barrel Ring Imaging CHerenkov (RICH).
The combination of the data from the TPC and RICH detectors and kinematic
cuts provide an efficient detection of new heavy particles with 
a small background for masses from 2 GeV/c$^2$ to the kinematic limit.

 The data taken during the period from 1995 to 1997 corresponds to an integrated luminosity 
of  11.9 pb$^{-1}$ at an energy of 130-136 GeV (including 6 pb$^{-1}$ taken in 
1997) 9.8 pb$^{-1}$ at an energy of 161 GeV,  9.9 pb$^{-1}$ at an energy of 172 GeV,
and 54.0 pb$^{-1}$ at an energy of 183 GeV.

\section{Event selection}

 A description of the DELPHI apparatus and its performance can be 
found in ref.\cite{ref:delp}, with more details on the Barrel RICH in 
ref. \cite{ref:brich} and
particle identification using the RICH in ref. \cite{ref:ribmean}.

 Charged particles were selected if their impact parameter with respect to the
mean beam-spot
was less than 5 cm in the 
$xy$ plane (perpendicular to the beam), and less than 10 cm in $z$ (the 
beam direction),
and their polar angle ($\theta$) was between 20 and 160 degrees.
The relative error on the measured momentum was required to be less than 1
and the track length larger than 30 cm. The energy of a charged particle was 
evaluated from its momentum\footnote{In the following, `momentum' means the apparent momentum, defined as 
the momentum divided by the charge $|q|$, because this is the physical quantity 
measured from the track curvature in the 1.23 T magnetic field.}
assuming the pion mass. Neutral particles were selected 
if their deposited energy was larger than 0.5 GeV and their polar angle 
was between 2 and 178 degrees.

The event was divided into two hemispheres using the thrust axis.
The total energy in one hemisphere was required to be larger than 10 GeV 
and the total energy of the charged particles in the other 
hemisphere to be larger than 10 GeV.
The event must have at least 
two reconstructed charged particle tracks including
at least one charged particle with momentum 
above 5 GeV/c reconstructed by the TPC and also
inside the acceptance of the Barrel RICH, $|\cos\theta|<0.68$.

 Cosmic muons were removed by putting tighter cuts on the impact parameter 
with respect to the mean beam-spot position.
When the event had two charged particles with at least one 
identified muon in the muon chambers, the impact parameter 
in the XY plane was required to be less than 0.15 cm, and below
1.5 cm in Z.

 The highest momentum (leading) charged particle in a given 
hemisphere was selected and identified 
using a combination of the following signals 
(where the typical sensitive mass range for pair produced 
sleptons at an energy of 183 GeV is 
shown in brackets):\\
 (1) the Gas Veto:  no photons were observed in the Gas RICH 
($m>$1 GeV/c$^2$) \\
 (2) the Liquid Veto:  four or less photons were observed in 
the Liquid RICH ($m>$65 GeV/c$^2$)\\
 (3) high ionization loss in the TPC: measured ionization 
was above 2 units i.e. twice the energy loss for a minimum ionizing particle 
($m>$70 GeV/c$^2$)\\
 (4) low ionization loss in the TPC: measured ionization
was below that expected for protons ($m$=1-50 GeV/c$^2$)\\
Selections (1) till (3) are identical to those used in our 
previous publication \cite{ref:hdelphi}.

 For the Gas and Liquid Vetoes it was required that the RICH was 
fully operational and that for a selected track 
photons from other tracks or ionization hits were detected 
inside the drift tube crossed by the track.
Due to tracking problems electrons often passed a Gas or Liquid Veto.
Therefore it was required that particles that deposit more than 5 GeV 
in the electromagnetic calorimeter, had either hits included in 
the outer tracking detector or associated RICH ionization hits.
At least 80 from a maximum of 160 wires were required for the measurement 
of the ionization in the TPC.

 Two sets of cuts selected sleptons 
or squarks. One set was defined for `leptonic topologies' for which the number 
of charged particles is less than four and another set for `hadronic topologies'
for all other events.
The cuts were optimized using slepton and squark events 
generated with SUSYGEN \cite{ref:stavros} and passed through 
the detector simulation program \cite{ref:delp}.
 Samples with different masses for smuons, free squarks with a charge of 
$\pm \frac{2}{3}e$ and hadronizing 
sbottom and stop squarks were studied in detail.

The hadronization of squarks was implemented in the following way.
The initial squark four-momenta including initial state 
radiation were generated by SUSYGEN. The JETSET parton shower model was 
used to fragment the squark-anti-squark string \cite{ref:jetset}.
In the fragmentation process 
the Peterson fragmentation function was used with a value for $\epsilon=
0.003\:(5/m)^2$, where $m$ is the mass of the squarks in GeV/c$^2$ 
\cite{ref:been}.
A shadron was given the mass of the squark plus 
150 MeV/c$^2$ for a smeson or plus 300 MeV/c$^2$ for a sbaryon.
In the fragmentation process, approximately 9\% sbaryons were formed and 40\%
of the shadrons were 
charged, 60 \% neutral. In the detector simulation program a charged shadron 
was given the properties of a heavy muon, a neutral shadron those of 
a K$^0_L$\footnote{It was only required that a charged shadron leaves a track as for a
  particle with unit charge, and that a neutral shadron deposit most of
  its energy in the hadron calorimeter.}. Due to the hard fragmentation function the charged multiplicity 
decreases as a function of the mass of the squark. At very high masses 
a squark-antisquark pair often produces a low multiplicity final state.

 For leptonic topologies an event was selected if the momentum of the charged 
particle was above 15 GeV/c and the Gas Veto (1) was confirmed by a Liquid Veto (2) or 
a low ionization loss (4) (in boolean notation (1)$\cdot$(2)+(1)$\cdot$(4)) or if 
the momentum of the charged particle 
was above 5 GeV/c and the Gas Veto was confirmed by a high ionization loss 
((1)$\cdot$(3)). The event was also accepted if both hemispheres had 
charged particles with momenta 
above 15 GeV/c and both leading charged particles had a Gas Veto or a high ionization 
loss or both a low ionization loss (((1)+(3))$\cdot$((1')+(3'))+(4)$\cdot$(4')),
where the primed selections refer to the opposite hemisphere.

 For hadronic topologies the following kinematic quantities were 
used to select events where a large fraction of the energy is taken by
a heavy particle. The energy fraction, $F_{c}$, is
defined as the momentum of the identified 
charged particle 
divided by the total energy in a given hemisphere, and $F_{n}$ the ratio of the 
neutral energy with respect to the total energy in a hemisphere.
The energy fraction $F$ is the maximum of $F_{c}$ and $F_{n}$.
The background from normal $q\bar{q}$ events 
was greatly reduced by requiring a minimum energy fraction $F$,
because heavy shadrons take most of the energy.

An event in a hadronic topology was selected if the momentum of the 
leading charged particle 
was above 15 GeV/c, the energy fraction $F$ was above 0.6 in one
hemisphere and above 0.9 in the other.
The selected charged particle had to be identified by a Gas Veto  or 
a high or a low ionization loss ((1)+(3)+(4)).

An event was also selected if the energy fraction $F$
in one hemisphere was above 0.6. In this case 
the momenta of the charged particles in both hemispheres had to be above 15 GeV/c and 
both leading charged particles had a Gas Veto, or both had high ionization,
or both low ionization  ((1)$\cdot$(1')+(3)$\cdot$(3')+(4)$\cdot$(4')).

\section{Analysis results}

No event was selected in the leptonic topology. The expected background 
was evaluated from the data and estimated to be 0.7 $\pm$ 0.3  events.
In Figure 1 the data taken at 183 GeV are shown for leptonic topologies.
The measured ionization and the measured Cherenkov angle 
in the liquid radiator are shown after applying the Gas Veto.

Three events were selected in the hadronic topology:
one at 130 GeV, one at 161 GeV and one at 183 GeV.
The expected background was estimated to be 3.5 $\pm$ 1.5 events 
using the real data and assuming that the 
background is from Standard Model processes, when the 
RICH or TPC misidentifies a particle known to be a pion (electron, muon, kaon or proton) as 
a heavy particle. The misidentification probability was evaluated from the 
data and used to estimate the expected background. The procedure 
was cross-checked by simulation studies.
The three candidate events have total charged multiplicities 
of  6, 4 and 5. The masses of the hypothetical squarks were estimated from a 
constrained fit using energy and momentum conservation
and found to be 48, 21 and 30 GeV/c$^2$ with typical uncertainties
of about $\pm$10 GeV/c$^2$. The mass is also correlated to 
the charged multiplicity. The most likely squarks masses
based both on these masses and the observed charged multiplicities are
41, 30 and 42 GeV/c$^2$. The resulting probability density
distribution is not very gaussian. The characteristics of the candidate events
are compatible with the background expectation.
Figure 2 shows the data taken at 183 GeV for hadronic 
topologies. The data are shown after the kinematic cut (see section 2)
requiring that the energy fraction $F$ was above 60\% in both 
hemispheres and in one of the hemispheres above 90\%.
One candidate event passes the Gas Veto (Fig. 2b).

The efficiency for selecting an event 
was evaluated as a function of the mass at different energies 
for right-handed smuons, mixed free stop quarks of charge 
q=$\pm \frac{2}{3}e$, mixed hadronizing stop quarks and 
mixed hadronizing sbottom quarks.
The term `mixed' refers to a typical mixing angle between left- and right-handed 
particles for which the cross-section is minimal. The angle 
is $\sim$60 degrees for stop quarks and $\sim$70 degrees for sbottom quarks.
The efficiency curves for a centre-of-mass energy of 183 GeV are shown 
in Figures 3a to 6a. The efficiency approaches zero at masses below
1 GeV/c$^2$, where the Gas Veto becomes inefficient. Therefore the lowest
upper limit on the mass is put at 2 GeV/c$^2$.

The efficiency curves for left- and right-handed squarks are slightly 
different due the different kinematical distributions,
but this difference can be neglected because it has no 
influence on the quoted upper limits.

 The efficiency curves have an overall systematic error 
of $\pm$5\% coming from the modelling of the detector.
For the hadronization of squarks the following effects 
were studied using the simulation:
a change in the fraction of neutral shadrons,
the response of the calorimeter to a neutral shadron and 
the fragmentation function.
In the simulation the fraction of neutral shadrons is 
60\%. This was changed to 50\% and an efficiency increase of 
15\% was found. In the simulation it was assumed that a neutral 
shadron behaves like a $K^0_L$. If one assumes that a neutral shadron 
deposits only 20\% of the energy of a $K^0_L$ and the rest escapes,
the efficiency is only reduced by 10\%. Finally the fragmentation function 
was softened assuming that $\epsilon$ is inversely proportional to the squark mass 
with $\epsilon=0.003\:(5/m)$. The efficiency at a centre-of-mass 
energy of 183 GeV increased by 20\% around a 
squark mass of 45 GeV/c$^2$ and decreased by 15\% around 70 GeV/c$^2$.
From these studies it was concluded that the efficiencies
for squarks are sufficiently stable under these large changes.

The observed numbers of events in the leptonic 
and hadronic topologies are compatible with the expected background.
Experimental upper limits at 95\% confidence level 
are obtained on the cross-section in the leptonic 
and hadronic topologies.  
In the leptonic topology the 95\% confidence level upper limit corresponds to 3 events.
In the hadronic topology it corresponds to 5.4 events in 
the case of 
3 observed events with 3 expected background events. 

The masses and charged particle multiplicity distributions of 
the candidates are included in the experimental upper limit.
From the simulation, the probability distribution 
as a function of the squark mass is obtained for each candidate and 
the sum of these 3 probability distributions is shown 
in Figure 7. 
The upper limit on the number of events
at 95\% confidence level 
is derived from this distribution by scaling it and adding it to 3.
Zero probability in this figure would thus correspond to an upper limit of 3 events.
The scale factor is adjusted such that  
3 observed events with a flat probability distribution 
would correspond to an upper limit of 5.4 events.
It was cheked that this procedure is  sufficiently precise
for the present analysis.
The experimental upper limit on the cross-section was derived
from the upper limit on the number of events, 
the signal efficiencies, integrated luminosities and cross-section ratios
at different energies as explained in footnote 6 of ref. \cite{ref:hdelphi}.

Figures 3 and 4 summarize the results for the leptonic topology for 
stable and long-lived sleptons, charginos and free squarks while Figures 5 and
6 summarize the results for the hadronic and leptonic 
topologies for stable and long-lived squarks.

Figure 3b shows the expected production cross-section for right- and left-handed 
smuons (staus) as a function of the mass at a centre-of-mass energy 
of 183 GeV. The combined experimental upper limit at 95\% confidence level 
on the cross-section varies between 0.06 and 0.5 pb in the mass range from 2 to 
90 GeV/c$^2$.
Right(left)-handed smuons or 
staus are excluded in their mass range from 2 to 80 (81) GeV/c$^2$.
From the same data, stable and long-lived charginos are excluded 
in the mass region from 2 to 87.5 GeV/c$^2$ for 
sneutrino masses above 41 GeV/c$^2$. For sneutrino masses 
above 200 GeV/c$^2$ the excluded mass goes up to 89.5 GeV/c$^2$.

Figure 4b shows the expected production cross-section for free 
mixed (right, left-handed) stop quarks as a function of the mass at an energy 
of 183 GeV. The combined experimental upper limit at 
95\% confidence level 
varies between 0.06 and 0.5 pb in the mass range from 2 to 
90 GeV/c$^2$.
Free mixed  (right, left-handed) stop quarks 
are excluded in the mass range from 2 to 84 (84, 86) GeV/c$^2$.
Similarly, free right(left)-handed 
up-type squarks of charge $\pm \frac{2}{3}e$
are excluded in the range from 2 to 84 (86) GeV/c$^2$.

Figure 5b shows the expected production cross-section for 
mixed (right, left-handed) stop quarks as a function of the mass 
at an energy of 183 GeV.
The combined experimental upper limit at 95\% confidence level on the 
cross-section varies between 0.1 and 0.5 pb in the mass range from 5 to 
90 GeV/c$^2$.
Hadronizing mixed  (right, left-handed) stop quarks are excluded in 
the mass range from 2 to 80 (81, 85) GeV/c$^2$.
Similarly, hadronizing right(left)-handed 
up-type squarks are excluded in the range from 2 to 81 (85) GeV/c$^2$.

Figure 6b shows the expected production cross-section for 
mixed (right, left-handed) sbottom quarks as a function of the mass 
at an energy of 183 GeV.
The combined experimental upper limit at 95\% confidence level on the 
cross-section is also shown. It varies between 0.15 and 0.5 pb in the mass range from 5 to 
90 GeV/c$^2$.
Hadronizing mixed (right, left-handed) sbottom quarks 
are excluded in the mass range from 5 (5, 2) to 38 (40, 83) GeV/c$^2$.
Similarly, right(left)-handed down-type squarks 
are excluded in the range from 5 (2) to 40 (83) GeV/c$^2$.

These results supersede those previously published \cite{ref:hdelphi}.

\section{Conclusions}

A search is made for stable and long-lived heavy charged particles 
in leptonic and hadronic final states 
at energies from 130 to 183 GeV, using particles identified 
by the Cherenkov light in the RICH and the 
ionization loss in the TPC.

No event is observed in the leptonic topology with an expected background of 
0.7 $\pm$ 0.3 events. In the hadronic topology 3 events were observed with an expected 
background of 3.5 $\pm$ 1.5 events.
The upper limit at 95\% confidence level 
on the cross-section at a centre-of-mass energy of 
183 GeV for sleptons 
and free squarks of charge $\pm \frac{2}{3}e$ varies between 0.06
and 0.5 pb in the mass range from 2 to 90 GeV/c$^2$.
The upper limit for hadronizing 
squarks varies between 0.15 and 0.5 pb in the mass range 
from 5 to 90 GeV/c$^2$.
Table 1 summarizes the excluded mass region at 95\% confidence level 
for different stable and long-lived supersymmetric particles.

\begin{table}[ht]
\begin{center}
\begin{tabular}{|c|c|}\hline
particle   &   excluded mass range \\
      &    $GeV/c^2$    \\ \hline \hline 
leptonic topologies &     \\ \hline 
$\tilde{\mu}_{R}$ or $\tilde{\tau}_{R}$ &  2-80  \\ \hline 
$\tilde{\mu}_{L}$ or $\tilde{\tau}_{L}$ &  2-81  \\ \hline 
$\tilde{\chi}^{\pm}$ ($m_{\tilde{\nu}}>41$ GeV/c$^2$) &  2-87.5  \\ \hline 
$\tilde{\chi}^{\pm}$ ($m_{\tilde{\nu}}>200$ GeV/c$^2$) &  2-89.5  \\ \hline 
free squarks       &      \\ \hline 
$\tilde{t}$ mixed &  2-84  \\ \hline 
$\tilde{t}_{R}$ or up-type $\tilde{q_{R}}$ &  2-84  \\ \hline 
$\tilde{t}_{L}$ or up-type $\tilde{q_{L}}$ &  2-86  \\ \hline\hline 
hadronic and leptonic topologies &     \\ \hline 
hadronizing squarks       &      \\ \hline 
$\tilde{t}$ mixed &  2-80  \\ \hline 
$\tilde{t}_{R}$ or up-type $\tilde{q_{R}}$ &  2-81  \\ \hline 
$\tilde{t}_{L}$ or up-type $\tilde{q_{L}}$ &  2-85  \\ \hline 
$\tilde{b}$ mixed &  5-38  \\ \hline 
$\tilde{b}_{R}$ or down-type $\tilde{q_{R}}$ &  5-40  \\ \hline 
$\tilde{b}_{L}$ or down-type $\tilde{q_{L}}$ &  2-83  \\ \hline\hline 
\end{tabular}
\caption{Excluded mass range at 95\% confidence level for stable and long-lived particles}
\end{center}
\end{table}








\pagebreak
\subsection*{Acknowledgements}
\vskip 3 mm
 We are greatly indebted to our technical 
collaborators, to the members of the CERN-SL Division for the excellent 
performance of the LEP collider, and to the funding agencies for their
support in building and operating the DELPHI detector.\\
We acknowledge in particular the support of \\
Austrian Federal Ministry of Science and Traffics, GZ 616.364/2-III/2a/98, \\
FNRS--FWO, Belgium,  \\
FINEP, CNPq, CAPES, FUJB and FAPERJ, Brazil, \\
Czech Ministry of Industry and Trade, GA CR 202/96/0450 and GA AVCR A1010521,\\
Danish Natural Research Council, \\
Commission of the European Communities (DG XII), \\
Direction des Sciences de la Mati$\grave{\mbox{\rm e}}$re, CEA, France, \\
Bundesministerium f$\ddot{\mbox{\rm u}}$r Bildung, Wissenschaft, Forschung 
und Technologie, Germany,\\
General Secretariat for Research and Technology, Greece, \\
National Science Foundation (NWO) and Foundation for Research on Matter (FOM),
The Netherlands, \\
Norwegian Research Council,  \\
State Committee for Scientific Research, Poland, 2P03B06015, 2P03B03311 and
SPUB/P03/178/98, \\
JNICT--Junta Nacional de Investiga\c{c}\~{a}o Cient\'{\i}fica 
e Tecnol$\acute{\mbox{\rm o}}$gica, Portugal, \\
Vedecka grantova agentura MS SR, Slovakia, Nr. 95/5195/134, \\
Ministry of Science and Technology of the Republic of Slovenia, \\
CICYT, Spain, AEN96--1661 and AEN96-1681,  \\
The Swedish Natural Science Research Council,      \\
Particle Physics and Astronomy Research Council, UK, \\
Department of Energy, USA, DE--FG02--94ER40817. \\

\begin{figure}[htb]
\begin{center}
\mbox{\epsfxsize18.0cm 
\epsffile{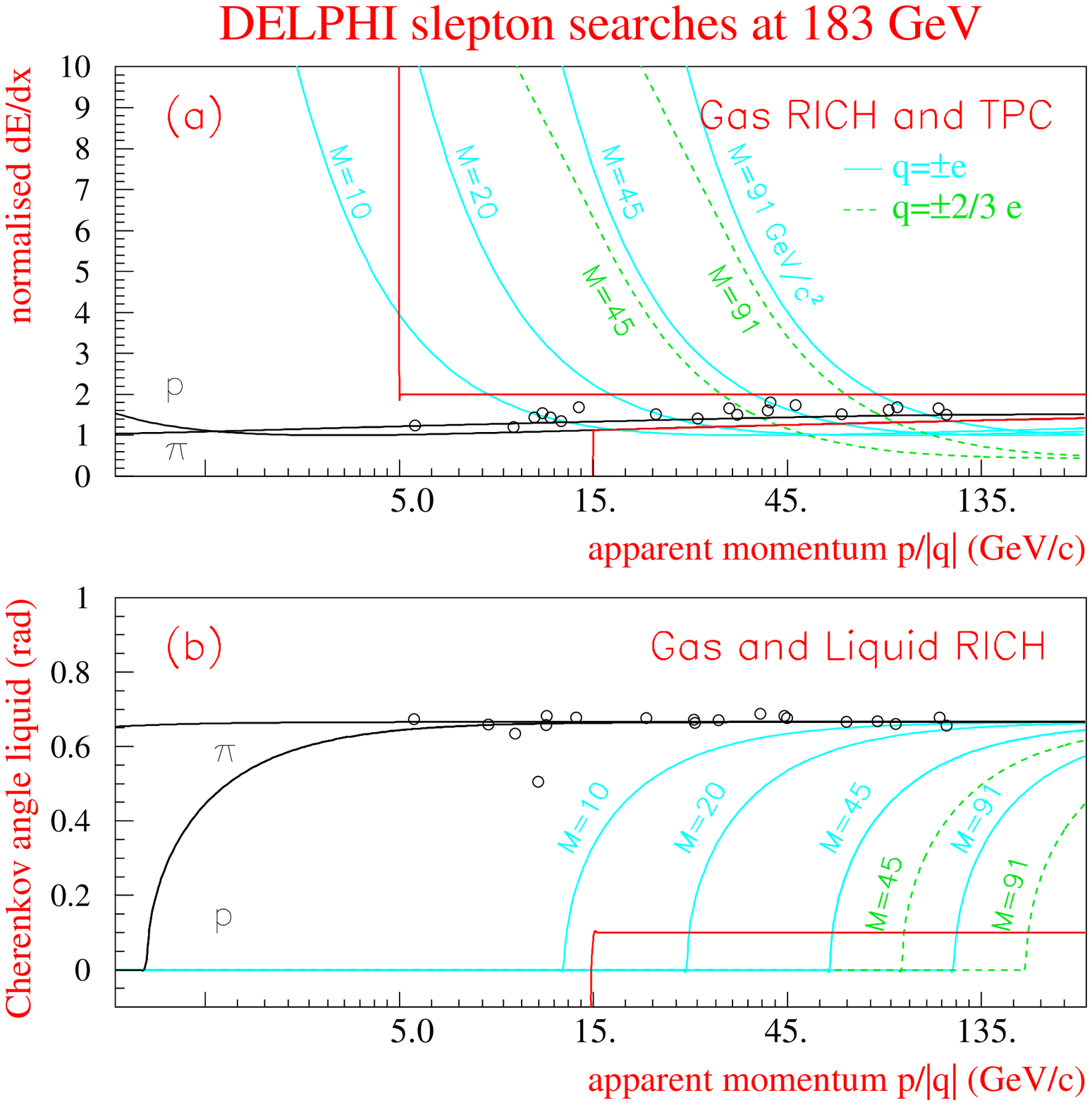} }
\end{center}
\vspace*{0.cm}
\caption{For leptonic topologies after the Gas Veto.
(a) Ionization as a function of the apparent momentum 
p/$|q|$ for the 183 GeV data.
(b) Measured Cherenkov angle in the liquid radiator 
as a function of the apparent momentum;
if four photons or less were observed 
in the liquid radiator, the Cherenkov angle was set equal to zero.
The expectation curves for charge $\pm e$ particles for pions, protons and 
heavy particles with masses of 10, 20, 45 and 91 GeV/c$^2$ are given, as well as 
the dashed curves for charge $\pm \frac{2}{3}e$ particles with masses of 45 and 91 GeV/c$^2$.
The areas bounded by straight lines in (a) indicate selections (3) and (4),
and that in (b) shows selection (2). The selection criteria are explained in section 2.}
\end{figure}

\begin{figure}[htb]
\begin{center}
\mbox{\epsfxsize18.0cm 
\epsffile{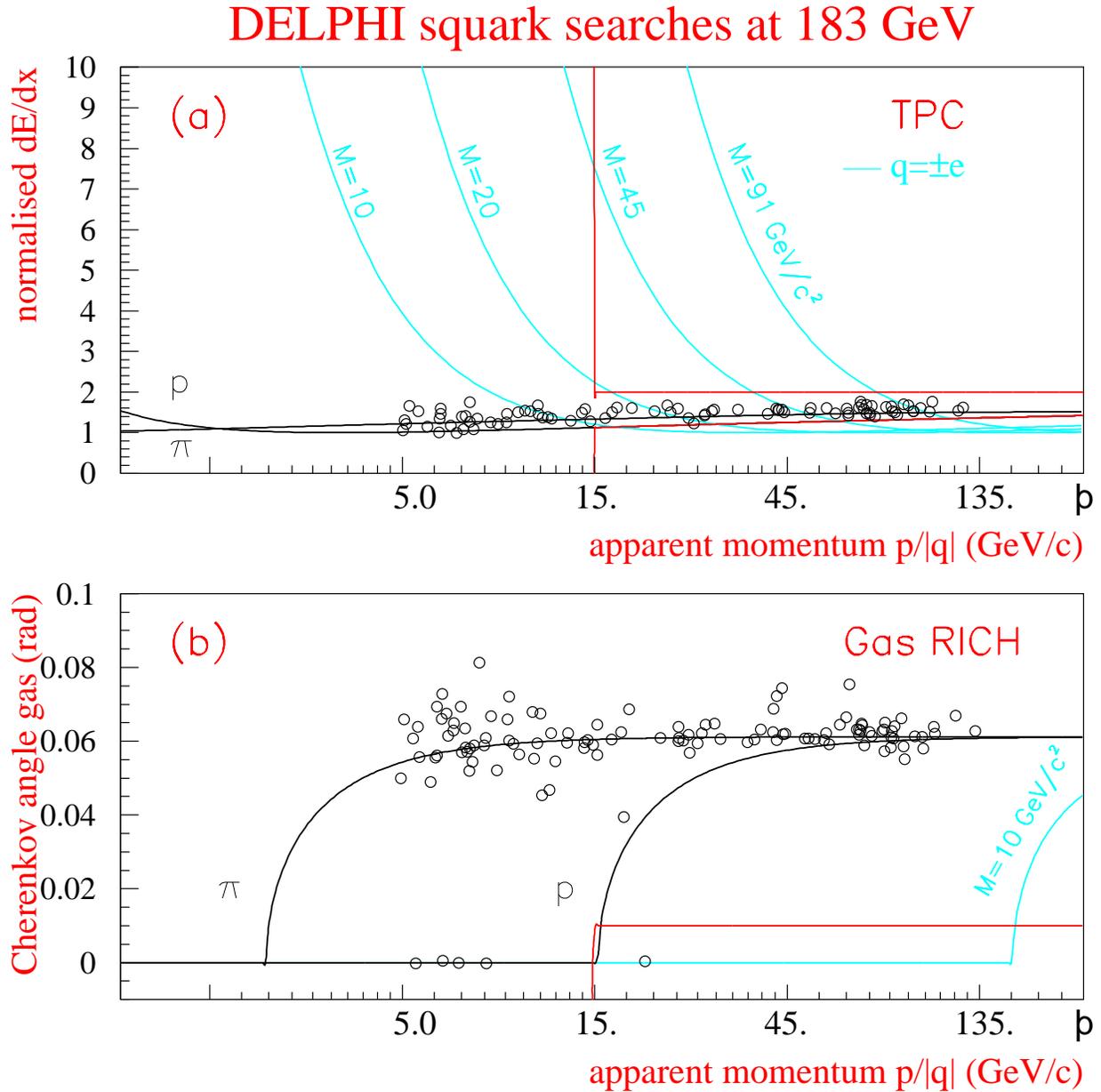} }
\end{center}
\vspace*{0.cm}
\caption{For hadronic topologies after the kinematic selection described in the 
text. (a) Ionization as a function of the apparent momentum 
p/$|q|$ for the 183 GeV data.
(b) Measured Cherenkov angle in the gas radiator 
as a function of the apparent momentum: if zero photons were observed 
the Cherenkov angle was set equal to zero.
The expectation curves for charge $\pm e$ particles for pions, protons and 
heavy particles with masses of 10, 20, 45 and 91 GeV/c$^2$ are given.
The areas bounded by straight lines in (a) indicate selections (3) and (4),
and that in (b) shows selection (1).
The selection criteria are explained in section 2. Only one candidate is in the 183 GeV data, the other two being at 
130 GeV and 161 GeV.}
\end{figure}

\begin{figure}[htb]
\begin{center}
\mbox{\epsfxsize18.0cm 
\epsffile{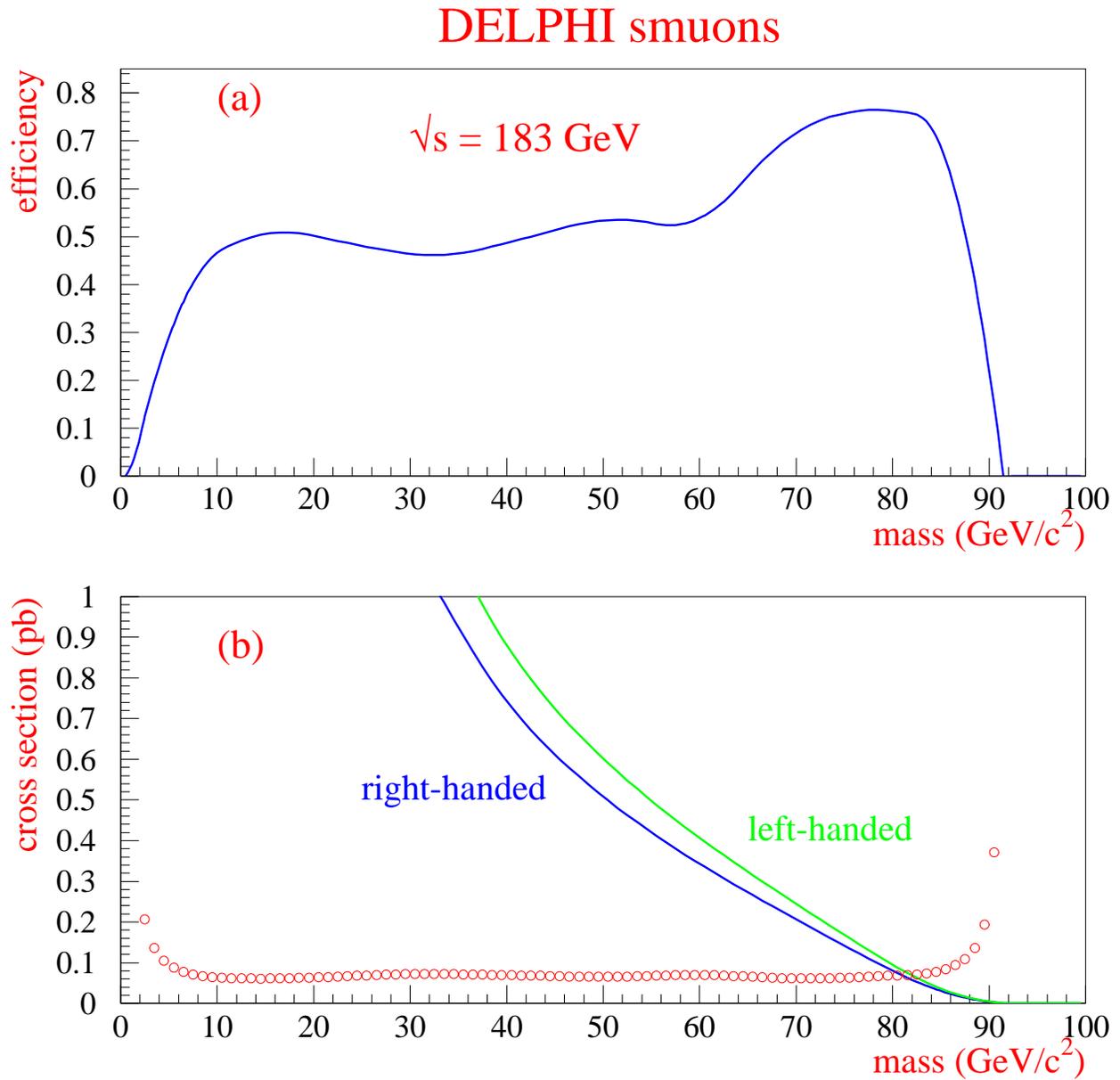} }
\end{center}
\vspace*{0.cm}
\caption{(a) Efficiency for detecting stable and long-lived smuons (staus)
as a function of the smuon mass at a centre-of-mass energy of 183 GeV.
(b) Production cross-section from SUSYGEN as a function of the smuon (stau) mass 
for right- and left-handed smuons at 183 GeV (full curves). The circles 
indicate the experimental 95\% confidence level upper limit for the combined 
130-136,161,172 and 183 GeV  data.}
\end{figure}

\begin{figure}[htb]
\begin{center}
\mbox{\epsfxsize18.0cm 
\epsffile{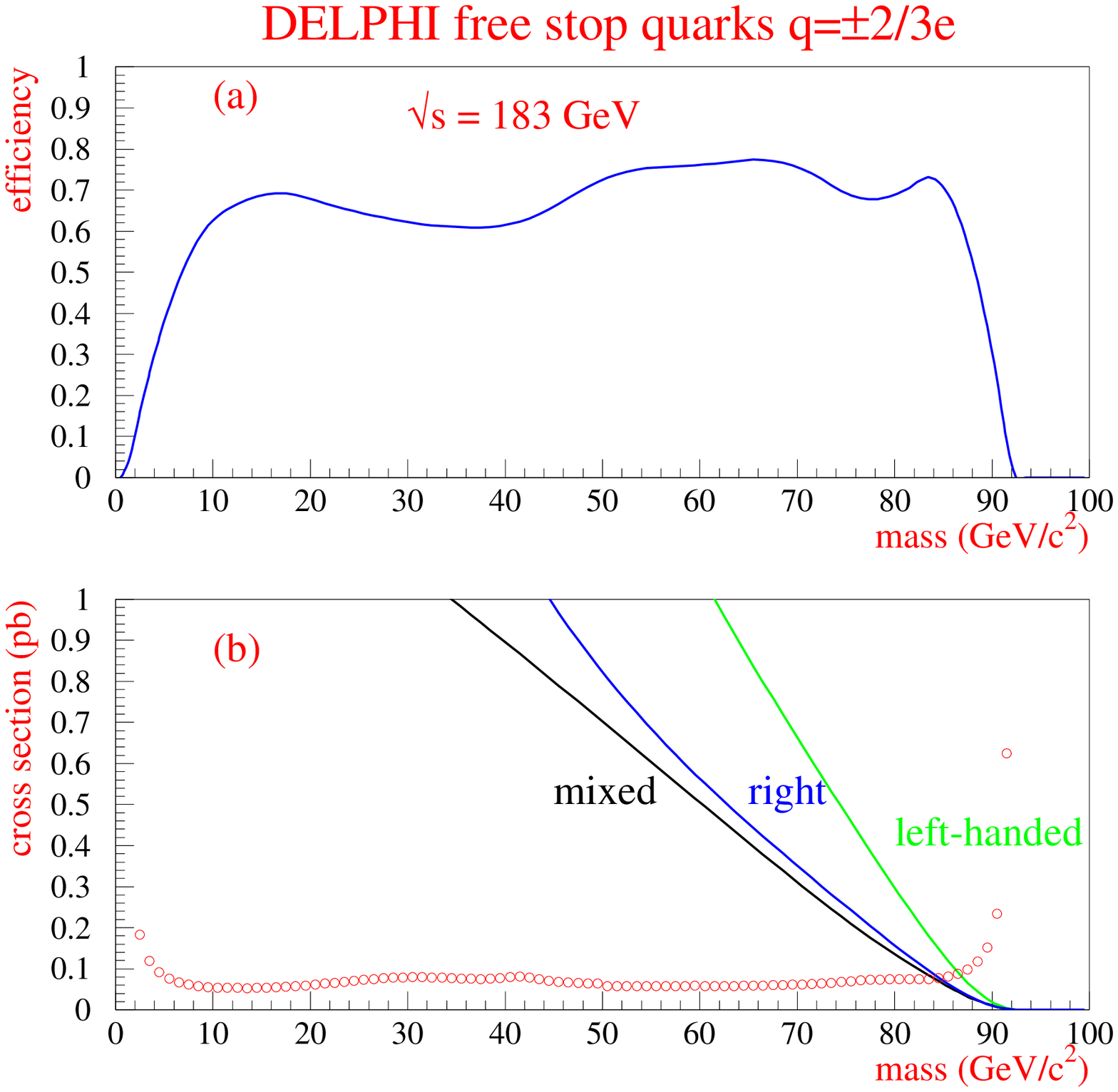} }
\end{center}
\vspace*{0.cm}
\caption{(a) Efficiency for detecting free stop quarks as a function of the stop mass 
at a centre-of-mass energy of 183 GeV.
(b) Production cross-section from SUSYGEN as a function of the stop mass 
for typical mixing, right- and left-handed stop quarks at 183 GeV 
(full curves). The circles 
indicate the experimental 95\% confidence level upper limit for the combined 
130-136,161,172 and 183 GeV  data.}
\end{figure}

\begin{figure}[htb]
\begin{center}
\mbox{\epsfxsize18.0cm 
\epsffile{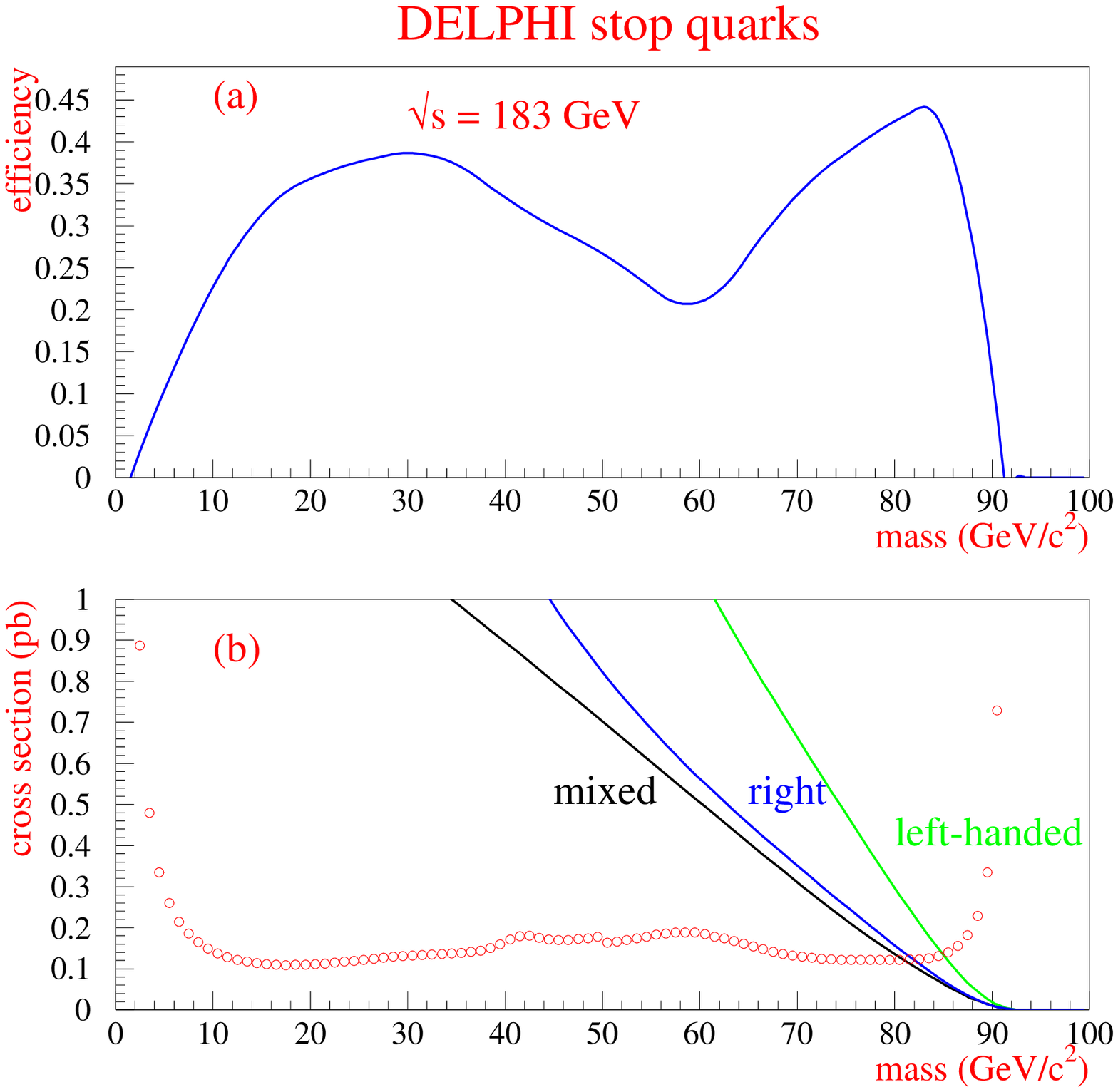} }
\end{center}
\vspace*{0.cm}
\caption{(a) Efficiency for detecting hadronizing stop quarks 
as a function of the stop mass 
at a centre-of-mass energy of 183 GeV.
(b) Production cross-section from SUSYGEN  as a function of the stop mass 
for typical mixing, right- and left-handed stop quarks at 183 GeV 
(full curves). The circles 
 indicated the experimental 95\% confidence level upper limit for the combined 
130-136,161,172 and 183 GeV  data.}
\end{figure}

\begin{figure}[htb]
\begin{center}
\mbox{\epsfxsize18.0cm 
\epsffile{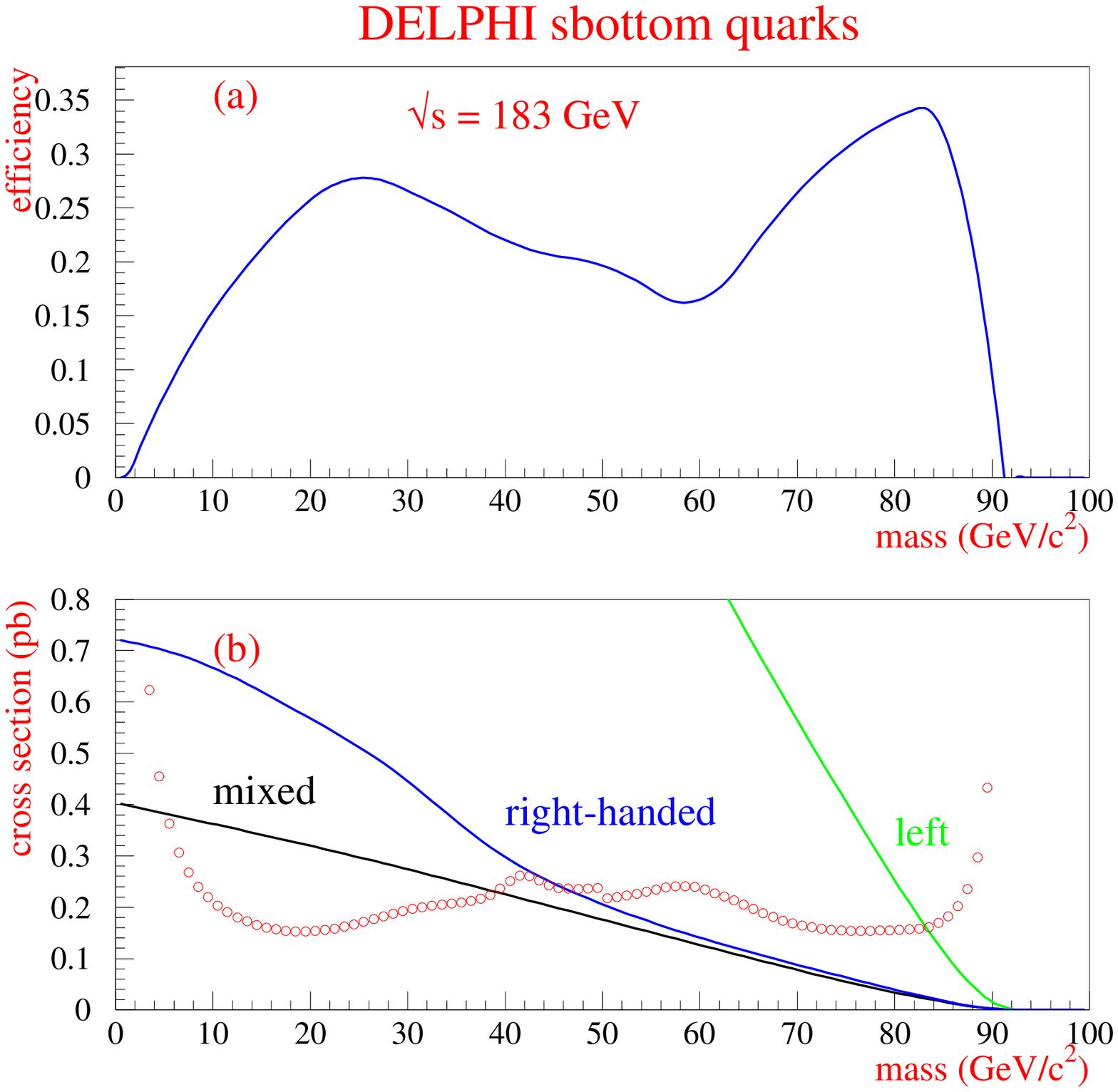} }
\end{center}
\vspace*{0.cm}
\caption{(a) Efficiency for detecting hadronizing sbottom quarks as a function of the 
sbottom mass at a centre-of-mass energy of 183 GeV.
(b) Production cross-section from SUSYGEN as a function of the sbottom mass 
for typical mixing, right- and left-handed sbottom quarks at 183 GeV 
(full curves). The circles 
 indicate the experimental 95\% confidence level upper limit for the combined 
130-136,161,172 and 183 GeV  data.}
\end{figure}

\begin{figure}[htb]
\begin{center}
\mbox{\epsfxsize18.0cm 
\epsffile{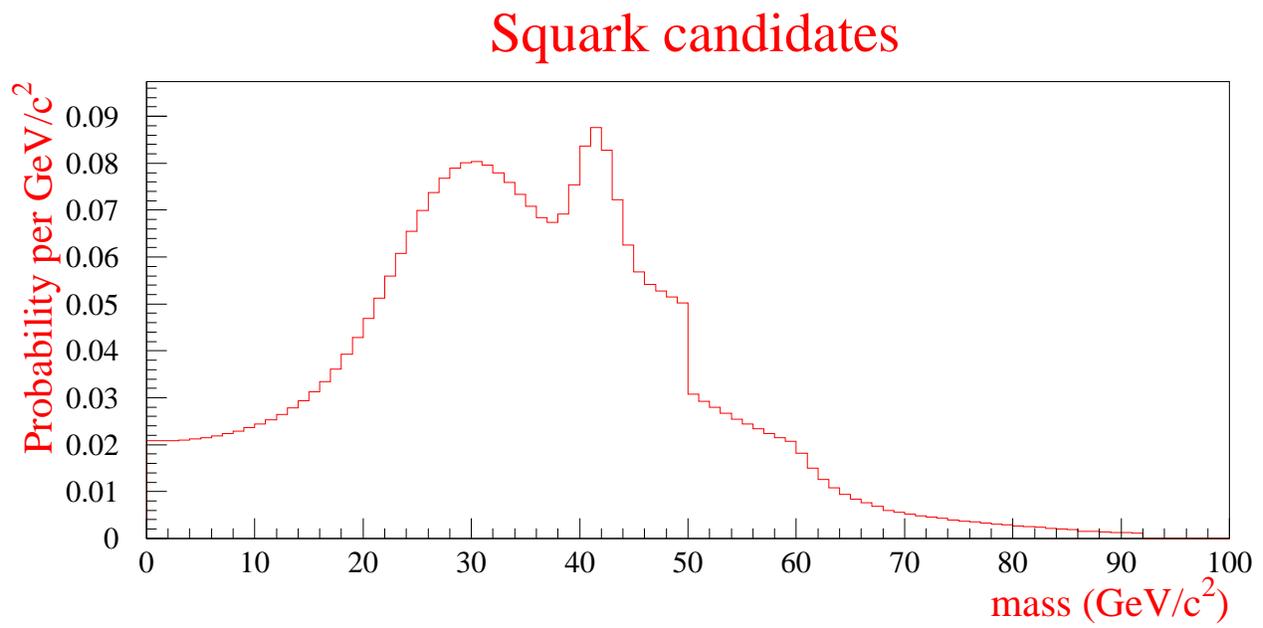} }
\end{center}
\vspace*{-7.5cm}
\caption{ Probability density distribution per GeV/c$^2$ 
for the three squark candidates (normalised to three) as a 
function of the squark mass.}
\end{figure}
\end{document}